\documentclass{ws-ijmpe}
\usepackage[super]{cite}
\usepackage{graphicx}
\usepackage[usenames]{color}
\usepackage[T1]{fontenc}
\newcommand{\bfr}{\mbox{\boldmath $r$}}

\newcommand{\bfx}{\mbox{\boldmath $x$}}

\newcommand{\slep}{\mbox{$\tilde{l}$}}
\newcommand{\stau}{\mbox{$\tilde{\tau}$}}
\newcommand{\grav}{\mbox{$\tilde{G}$}}

\begin{document}

\markboth{M. Kusakabe, G. J. Mathews, T. Kajino, M.-K. Cheoun}
{EFFECTS OF LONG-LIVED NEGATIVELY CHARGED MASSIVE PARTICLES}

\catchline{}{}{}{}{}

\title{REVIEW ON EFFECTS OF LONG-LIVED NEGATIVELY CHARGED MASSIVE PARTICLES ON BIG BANG NUCLEOSYNTHESIS}

\author{\footnotesize Motohiko Kusakabe\footnote{JSPS Postdoctoral Fellow for Research Abroad}, Grant. J. Mathews}

\address{Center for Astrophysics, Department of Physics, University of Notre Dame, Notre Dame, IN 46556, U.S.A. \\
mkusakab@nd.edu}

\author{Toshitaka Kajino}
\address{National Astronomical Observatory of Japan, 2-21-1 Osawa, Mitaka, Tokyo 181-8588, Japan \\
  Department of Astronomy, Graduate School of Science, University of Tokyo, 7-3-1 Hongo, Bunkyo-ku, Tokyo 113-0033, Japan}

\author{Myung-Ki Cheoun}
\address{Department of Physics, Soongsil University, Seoul 156-743, Korea}



\maketitle

\begin{history}
\received{4 July 2015}
\revised{12 April 2016}
\end{history}

\begin{abstract}
We review important reactions in the big bang nucleosynthesis (BBN) model involving a long-lived negatively charged massive particle, $X^-$, which is much heavier than nucleons.  This model can explain the observed $^7$Li abundances of metal-poor stars, and predicts a primordial $^9$Be abundance that is larger than the standard BBN prediction.  
In the BBN epoch, nuclei recombine with the $X^-$ particle.  Because of the heavy $X^-$ mass, the atomic size of bound states $A_X$ is as small as the nuclear size.  The nonresonant recombination rates are then dominated by the $d$-wave $\rightarrow$ 2P transition for $^7$Li and $^{7,9}$Be.  The $^7$Be destruction occurs via a recombination with the $X^-$ followed by a proton capture, and the primordial $^7$Li abundance is reduced.  Also, the $^9$Be production occurs via the recombination of $^7$Li and $X^-$ followed by deuteron capture.  The initial abundance and the lifetime of the $X^-$ particles are constrained from a BBN reaction network calculation.  
We estimate that the derived parameter region for the $^7$Li reduction is allowed in supersymmetric or Kaluza-Klein (KK) models.  We find that either the selectron, smuon, KK electron or KK muon could be candidates for the $X^-$ with $m_X\sim {\mathcal O}(1)$ TeV, while the stau and KK tau cannot.

\end{abstract}

\ccode{PACS Nos.: 26.35.+c, 95.35.+d, 98.80.Cq, 98.80.Es}

\section{INTRODUCTION}\label{sec1}
The primordial light element abundances calculated in standard big bang nucleosynthesis (SBBN) are more or less consistent with those inferred from astronomical observations.  A large discrepancy, however, exists in the primordial $^7$Li abundance.  Spectroscopic measurements of metal-poor stars (MPSs) indicates a roughly constant abundance ratio, $^7$Li/H$=(1-2) \times 10^{-10}$, as a
function of metallicity.\cite{Spite:1982dd,Ryan:1999vr,Melendez:2004ni,Asplund:2005yt,bon2007,shi2007,Aoki:2009ce,Hernandez:2009gn,Sbordone:2010zi,Monaco:2010mm,Monaco:2011sd,Mucciarelli:2011ts,Aoki:2012wb,Aoki2012b}  However, the theoretical primordial abundance in the SBBN
model is larger than the observational value by about a factor of $3-4$ (e.g., Refs. \refcite{Coc:2011az,Coc:2013eea}) when the baryon-to-photon ratio in the $\Lambda$CDM model is taken from observations of the cosmic microwave background radiation by the Wilkinson Microwave Anisotropy Probe\cite{Spergel:2003cb,Spergel:2006hy,Larson:2010gs,Hinshaw:2012fq} or the Planck observatory\cite{Ade:2013zuv}).  This discrepancy suggests that unknown physics is present to reduce the $^7$Li abundance during or after BBN.  The early Galaxy might have had a $^7$Li abundance smaller than the cosmic average value because of a chemical separation induced by the primordial magnetic field\cite{Kusakabe:2014dta}.  Rotationally induced mixing\cite{Pinsonneault:1998nf,Pinsonneault:2001ub}, and a combination of atomic and turbulent diffusion\cite{Richard:2004pj,Korn:2007cx,Lind:2009ta} might have reduced the $^7$Li abundance in stellar atmospheres.  As another possibility, an exotic particle might have existed during big bang nucleosynthesis (BBN), and affected the primordial abundances.

A late-decaying negatively charged massive particle (CHAMPs or Cahn-Glashow particles) $X^-$ has been considered \cite{cahn:1981,Dimopoulos:1989hk,rujula90} as a solution to the Li problem.\cite{Pospelov:2006sc,Kohri:2006cn,Cyburt:2006uv,Hamaguchi:2007mp,Bird:2007ge,Kusakabe:2007fu,Kusakabe:2007fv,Jedamzik:2007cp,Jedamzik:2007qk,Kamimura:2008fx,Pospelov:2007js,Kawasaki:2007xb,Jittoh:2007fr,Jittoh:2008eq,Jittoh:2010wh,Pospelov:2008ta,Khlopov:2007ic,Kawasaki:2008qe,Bailly:2008yy,Jedamzik:2009uy,Kamimura2010,Kusakabe:2010cb,Pospelov:2010hj,Kohri:2012gc,Cyburt:2012kp,Dapo2012,Kusakabe:2013tra,2013PhRvD..88h9904K,Kusakabe:2014moa}.  Constraints on supersymmetric $X^-$ models have also been derived through  BBN calculations.\cite{Cyburt:2006uv,Kawasaki:2007xb,Jittoh:2007fr,Jittoh:2008eq,Jittoh:2010wh,Pradler:2007ar,Pradler:2007is,Kawasaki:2008qe,Bailly:2008yy}  Long-lived CHAMPs have been searched for in collider experiments.  No signature of an $X^-$ has been observed, and limits on the mass have been placed by measurements at the Large Hadron Collider.  Lower limits on the mass of the scalar $\tau$ leptons (staus) are typically several hundred GeV, and depend on parameters of particle models.\cite{Chatrchyan:2013oca,CMS:2012xi,Aad:2012pra,ATLAS:2014fka,Aaij:2015ica}

The $X^-$ particles and nuclei $A$ can form bound atomic systems ($A_X$ or $X$-nuclei) with binding energies $\sim O(0.1-1)$~MeV in the limit that the mass of $X^-$, $m_X$, is much larger than the nucleon mass.\cite{cahn:1981,Kusakabe:2007fv}  The $X$-nuclei are exotic chemical species with very heavy masses and chemical properties similar to normal atoms and ions.  The present existence of the superheavy stable (long-lived) particles have been searched for in experiments.

Table \ref{ta1} shows constraints on the abundances of $X$-nuclei derived from experiment.  The first column shows upper limits on the number ratio of $X^-$ to nucleons.  The second column shows the element composing the sample used in the experiment, and the third column shows the mass region where the derived constraint is applicable.  The fourth column shows the reference for the experiment.

\begin{table}[ht]
\tbl{Experimental constraints on CHAMP.}
{\begin{tabular}{@{}cccc@{}} \toprule
$X/N$ & sample element & mass of $X^-$ & Ref. \\
\colrule
$<10^{-28}-10^{-29}$ & H & $m_X=11-1100$ GeV & \cite{Smith1982} \\
$<4\times 10^{-17}$ & H & $m_X=5-1500$ GeV & \cite{Yamagata:1993jq} \\
$<6\times 10^{-15}$ & H & $m_X=10^4-10^7$ GeV & \cite{Verkerk:1991jf} \\
$<5\times 10^{-12}$ & Na & $m_X=10^2-10^5$ GeV & \cite{Dick:1985wk} \\
$<2\times 10^{-15}$ & C & $m_X\leq 10^5$ GeV & \cite{Tur1984} \\
$<1.5\times 10^{-13}$ & Pb & $m_X\leq 10^5$ GeV & \cite{Norman:1988fd} \\ \botrule
$<2\times 10^{-24}-3\times 10^{-20}$ & H & & \\
$<2\times 10^{-13}-7\times 10^{-10}$ & Li & & \\
$<4\times 10^{-12}-7\times 10^{-9}$ & Be & & \\
$<4\times 10^{-14}-8\times 10^{-11}$ & B & $m_X= 10^2 -10^4$ GeV & \cite{Hemmick:1989ns} \\
$<4\times 10^{-20}-2\times 10^{-16}$ & C & & \\
$<4\times 10^{-17}-3\times 10^{-14}$ & O & & \\
$<7\times 10^{-15}-2\times 10^{-12}$ & F & & \\ \botrule
\end{tabular}\label{ta1} }
\end{table}

If the $X^-$ particle exists during the BBN epoch, it binds with nuclei and triggers nuclear reactions.\cite{Pospelov:2006sc,Kohri:2006cn,Cyburt:2006uv,Hamaguchi:2007mp,Bird:2007ge,Kusakabe:2007fu,Kusakabe:2007fv,Jedamzik:2007qk,Jedamzik:2007cp,Kamimura:2008fx,Pospelov:2007js,Kawasaki:2007xb,Jittoh:2007fr,Jittoh:2008eq,Jittoh:2010wh,Pospelov:2008ta,Khlopov:2007ic,Kawasaki:2008qe,Bailly:2008yy,Kamimura2010,Kusakabe:2010cb,Pospelov:2010hj,Kohri:2012gc,Cyburt:2012kp,Dapo2012,Kusakabe:2013tra,2013PhRvD..88h9904K,Kusakabe:2014moa}  The formation of most $X$-nuclei proceeds through radiative recombination \cite{Dimopoulos:1989hk,rujula90}of nuclides $A$ and $X^-$.  Importantly, the $^7$Be$_X$ formation proceeds also through the non-radiative $^7$Be charge exchange reaction \cite{Kusakabe:2013tra,2013PhRvD..88h9904K} between a $^7$Be$^{3+}$ ion and an $X^-$.

The $^6$Li abundance can significantly increase through the $X^-$-catalyzed transfer reaction $^4$He$_X$($d$, $X^-$)$^6$Li,\cite{Pospelov:2006sc} where 1(2$,$3)4 signifies a reaction $1+2\rightarrow 3+4$.

A relatively weak destruction of $^7$Be can also proceed, and the primordial $^7$Li abundance~\footnote{$^7$Be produced during the BBN is transformed into $^7$Li by electron capture in the epoch of the recombination of $^7$Be and electron much later than the BBN epoch.  The primordial $^7$Li abundance is, therefore, the sum of abundances of $^7$Li and $^7$Be produced in BBN.  In SBBN with the baryon-to-photon ratio inferred from the Planck\cite{Ade:2013zuv}, $^7$Li is produced predominantly as $^7$Be during the BBN.} is reduced.\cite{Bird:2007ge,Kusakabe:2007fu}  Bound states of $^8$B and $X^-$ include atomic states composed of nuclear ground and excited states of $^8$B.  In the reaction $^7$Be$_X$($p$,$\gamma$)$^8$B$_X$, the first atomic excited state of $^8$B$_X$,\cite{Bird:2007ge} and the atomic ground state of $^8$B$^\ast$($1^+$,0.770~MeV)$_X$ consisting of the $1^+$ nuclear excited state of $^8$B and an $X^-$,\cite{Kusakabe:2007fu} work as resonances that reduce $^7$Be abundance.

Effects on other nuclear abundances have been investigated in a large reaction network calculation for typical values of the $X^-$ abundance.\cite{Kusakabe:2007fu}  Especially, productions of nuclei with mass number $A\geq 9$ via several reactions including $^8$Be$_X$+$p$ $\rightarrow ^9$B$_X^{\ast{\rm a}} \rightarrow ^9$B$_X$+$\gamma$ through the $^9$B$_X^{\ast{\rm a}}$ atomic excited state of $^9$B$_X$ were studied.  No effect was, however, found in the abundances of nuclei with $A\geq 9$.\cite{Kusakabe:2007fv}

The resonant reaction
$^8$Be$_X$($n$, $X^-$)$^9$Be through the atomic ground state of
$^9$Be$^\ast$($1/2^+$, 1.684~MeV)$_X$\cite{Pospelov:2007js} has been found to be nonexistent since the state $^9$Be$^*$($1/2^+$, 1.684~MeV)$_X$ is not a
resonance but a bound state located below the $^8$Be$_X$+$n$ separation channel.\cite{Kamimura:2008fx}  This has been confirmed by a four-body calculation
for an $\alpha+\alpha+n+X^-$ system\cite{Kamimura2010} and another three-body calculation\cite{Cyburt:2012kp}.

The most important reaction of $^9$Be production has been found to be $^7$Li$_X$($d$, $X^-$)$^9$Be.\cite{Kusakabe:2014moa}  This reaction is the key reaction since a signature of the $X^-$ particle is left on primordial $^9$Be abundance through this reaction.  This model can, therefore, be tested by observations of $^9$Be abundances in MPSs in the future.  Therefore, realistic calculations of the reaction rate with quantum many-body models are needed.

In this paper we review the BBN model with a CHAMP, $X^-$ based upon our recent extensive study.\cite{Kusakabe:2014moa}  We then estimate the possibility that the $X^-$ particle which affects the primordial $^7$Li abundance corresponds to particles in supersymmetric or extra-dimensional models.  In Sec. \ref{sec2}, the adopted model for the nuclear charge density and the calculation of binding energies of the $X$-nuclei are explained.  In Sec.~\ref{sec3}, the radiative recombination of light nuclides with the $X^-$ particles are reviewed.  In Sec.~\ref{sec4}, a calculation of the radiative proton capture reactions $^{7,8}$Be$_X$($p$, $\gamma$)$^{8,9}$B$_X$ is shown.  In Sec.~\ref{sec5}, a reaction for $^9$Be production is described.  In Sec.~\ref{sec6}, our reaction network calculation is described.  In Sec.~\ref{sec7}, the latest observational constraints on the primordial nuclear abundances are described.  In Sec.~\ref{sec8}, we show the evolution of elemental abundances as a function of cosmic temperature, and derive updated constraints on the initial abundance and the lifetime of the $X^-$.  
In Sec.~\ref{sec9}, constraints on the $X^-$ particle for the reduction of the Li abundance are discussed.  In Sec. \ref{sec10} we summarize this review.

Throughout the paper, we use natural units, $\hbar=c=k_{\rm B}=1$, for the reduced Planck constant $\hbar$, the speed of light $c$, and the Boltzmann constant $k_{\rm B}$.  We use the usual notation 1(2$,$3)4 for a reaction $1+2\rightarrow 3+4$.

\section{MODEL}\label{sec2}
\subsection{NUCLEAR CHARGE DENSITY}\label{sec2a}

We assume that a CHAMP with a single negative charge and spin zero, $X^-$,  exists during the BBN epoch.  The mass of the $X^-$, i.e., $m_X$ is set to be $m_X=1000$ GeV.

In this review, we take the standard case of Ref. \refcite{Kusakabe:2014moa}.  It is assumed that the nuclear charge density is given by a Woods-Saxon shape,
\begin{equation}
 \rho_{\rm WS}(r^\prime)=\frac{ZeC_{\rm WS}}{1+\exp\left[\left(r^\prime-R\right)/a\right]},
\label{eq1}
\end{equation}
where
$r^\prime$ is the distance from the center of mass of the nucleus,
$Ze$ is the charge of the nucleus,
$R$ is the parameter characterizing the nuclear size,
$a$ is nuclear surface  diffuseness,
and $C_{\rm WS}$ is a normalization constant.
The $C_{\rm WS}$ value is fixed by the equation of charge conservation, $Ze=\int \rho_{\rm WS}~d\bfr^\prime$, and it is given by
\begin{equation}
 C_{\rm WS}=\left(4\pi \int_0^\infty \frac{{r^\prime}^2}{1+\exp\left[\left(r^\prime-R\right)/a\right]} dr^\prime\right)^{-1}.
\label{eq2}
\end{equation}
For a given value of diffuseness $a$, the size parameter $R$ can be constrained so that the parameter set of ($a$, $R$) satisfies the  root-mean-square (RMS) charge radius $\langle r^2 \rangle_{\rm C}^{1/2}$ measured in nuclear experiments.  The value of $a=0.40$ fm is chosen.

The potential between an $X^-$ and a nucleus $A$ ($XA$ potential) is calculated by folding the Coulomb potential with the charge density:
\begin{equation}
 V(r)=\int -\frac{e \rho_{\rm WS}(\bfr^\prime)}{x}~d\bfr^\prime,
\label{eq3}
\end{equation}
where
$\bfr$ is the position vector from an $X^-$ to the center of mass of $A$,
$\bfr^\prime$ is the position vector from the center of mass of $A$,
$\bfx=\bfr+\bfr^\prime$ is the displacement vector between the $X^-$ and the position, and $\rho_{\rm WS}(\bfr^\prime)$ is the charge density of the nucleus.  Under the assumption of a Woods-Saxon charge distribution $\rho_{\rm WS}(r^\prime)$, the potential reduces to the form
\begin{eqnarray}
 V_{\rm WS}(r)&=&-\frac{2\pi C_{\rm WS} Z e^2}{r} \int_0^\infty dr^\prime r^\prime
 \frac{(r+r^\prime)-|r-r^\prime|}{1+\exp[(r^\prime-R)/a]}~~.\nonumber\\
\label{eq4}
\end{eqnarray}

See Ref. \refcite{Kusakabe:2014moa} for a discussion of the dependences of binding energies, reaction rates, and BBN on $m_X$, the nuclear charge density, and experimental uncertainties in the RMS charge radii.
  
\subsection{BINDING ENERGY}\label{sec2b}

Binding energies and wave functions for bound states of nuclei $A+X^-$, i.e., $X$-nuclei or $A_X$, have been calculated\cite{Kusakabe:2014moa} using both numerical integrations of the Schr$\ddot{\rm o}$dinger equation with RADCAP\cite{Bertulani:2003kr} and variational calculations with the Gaussian expansion method\cite{Hiyama:2003cu}.

The reduced mass of the $A+X^-$ system is given by
\begin{equation}
  \mu=\frac{m_A m_X}{m_A+m_X},
  \label{eq_add1}
\end{equation}
where $m_A$ is the mass of nuclide $A$.  
In the limit of a heavy $X^-$ particle, i.e,. $m_X \gg m_A$, $\mu \rightarrow m_A$.  The binding energies, therefore, become independent of $m_X$.

\section{RADIATIVE RECOMBINATION WITH $X^-$}\label{sec3}

The recombination reactions of $^7$Be, $^7$Li, $^9$Be, and $^4$He with $X^-$ are important particularly, as seen in the BBN network calculation of Sec. \ref{sec8}.  The recombination rates for these four nuclides are therefore, precisely calculated.\cite{Kusakabe:2014moa}  Rates for other nuclides are, on the other hand, approximately given (Sec. \ref{sec3e}).

\subsection{$^7$Be}\label{sec3a}

\subsubsection{Energy Levels}\label{sec3a1}

Binding energies of $^7$Be$_X$ atomic states with main quantum numbers $n$ ranging from one to seven have been calculated.\cite{Kusakabe:2014moa}  Binding energies in the case of two point charges are given analytically by $E_{\rm B}^{\rm Bohr}=-Z^2 \alpha^2 \mu/2n^2$, where $\alpha$ is the fine structure constant.  Since the $^7$Be nuclear charge distribution has a finite size, the amplitude of the Coulomb potential at small $r$ is less than that for two point charges.  Wave functions at small radii and binding energies of tightly bound states with small $n$ values, therefore, deviate from those of the Bohr model.  On the other hand, the binding energies of loosely bound states with large $n$ values are similar to those of the Bohr model.

\begin{figure}
\begin{center}
\includegraphics[width=0.8\textwidth]{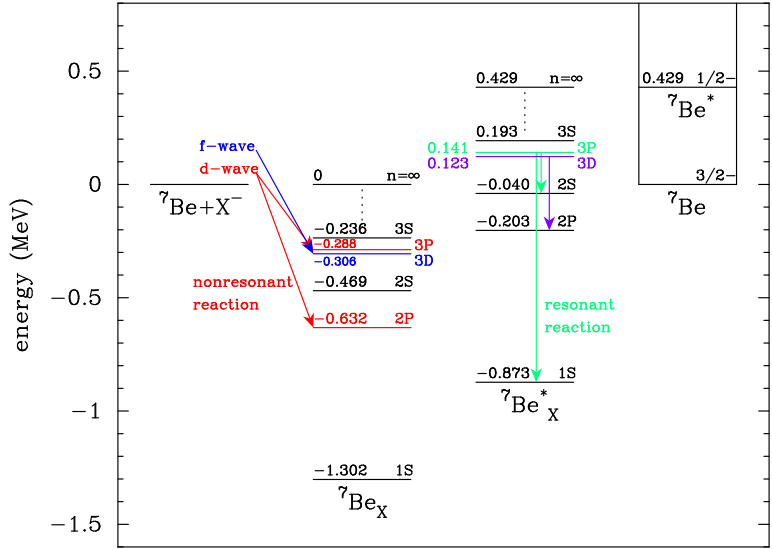}
\end{center}
\caption{Energy level diagram of $^7$Be$_X$ in the case of $m_X=1000$ GeV.  Red and blue arrows show the three important transitions in the nonresonant recombination reaction of $^7$Be($X^-$, $\gamma$)$^7$Be$_X$, while purple and green arrows show decays of the two important resonances into bound states of $^7$Be$^\ast_X$ in the resonant reaction.  \label{fig_new1}}
\end{figure}

Figure \ref{fig_new1} shows the energy level diagram of $^7$Be$_X$ in the case of $m_X=1000$ GeV.  The latest calculation indicates that the nonresonant rate of the recombination of $^7$Be and  $X^-$ is larger than the resonant rate.\cite{Kusakabe:2014moa}  Red and blue arrows show the three important transitions in the nonresonant recombination reaction (Sec. \ref{sec3a2}), while purple and green arrows show decays of the two important resonances into bound states of $^7$Be$^\ast_X$ in the resonant reaction (Sec. \ref{sec3a1}).


\subsubsection{$^7$Be($X^-$, $\gamma$)$^7$Be$_X$ Resonant Rate}\label{sec3a2}

The resonant rate of the reaction $^7$Be($X^-$, $\gamma$)$^7$Be$_X$ has been calculated taking into account the change of the E1 effective charge as a function of $m_X$.  The E1 effective charge is given by
\begin{equation}
e_1= e \frac{Z_1 m_2-Z_2 m_1}{m_1+m_2},
\label{eq16}
\end{equation}
where
$m_i$ and $Z_i$ are the mass and the charge number of species $i=1$ and 2.

The recombination of $^7$Be and $^7$Li with $X^-$ proceed via resonant reactions through atomic states ${^7A^\ast}_X^{\ast{\rm a}}$, composed of a nuclear excited state $^7A^\ast$ and an $X^-$.\cite{Bird:2007ge}  There are an infinite number of atomic states of $^7$Be$^\ast_X$, composed of the first nuclear excited state $^7$Be$^\ast$[$\equiv ^7$Be$^\ast$(0.429 MeV, $1/2^-$)] and an $X^-$.  However, the resonant reaction rate is suppressed by a factor of $\exp(-E_{\rm r}/T)$ with $E_{\rm r}$ the resonant energy [cf. Eq. (\ref{eq11})].  Therefore, only resonances whose energy levels are close to that of the entrance channel $^7$Be+$X^-$ are important.  In the case of $m_X=1000$ GeV, important resonances are then the 3P and the 3D states.

The thermal reaction rate is derived as a function of temperature $T$ by numerically integrating the cross section over energy,
\begin{equation}
\langle \sigma v \rangle =\left(\frac{8}{\pi \mu}\right)^{1/2} \frac{1}{T^{3/2}} \int_0^\infty E \sigma(E) \exp\left(-\frac{E}{T}\right) dE,
\label{eq10}
\end{equation}
where
$E$ is the center of mass kinetic energy,
and $\sigma(E)$ is the reaction cross section as a function of $E$.

The resonant rate is derived \cite{Kusakabe:2014moa} to be
\begin{eqnarray}
N_{\rm A} \langle \sigma v \rangle_{\rm R} &=& 
3.86 \times 10^4~{\rm cm}^3 {\rm mol}^{-1} {\rm s}^{-1} T_9^{-3/2} \exp(-1.43/T_9)  \nonumber \\
&&+1.44 \times 10^4~{\rm cm}^3 {\rm mol}^{-1} {\rm s}^{-1} T_9^{-3/2} \exp(-1.64/T_9). \label{eq23}
\end{eqnarray}
The first term corresponds to the atomic transition from the resonance $^7$Be${^\ast_X}^{\ast{\rm a}}$(3D) to $^7$Be${^\ast_X}^{\ast{\rm a}}$(2P), while the second term corresponds to sums of the atomic transitions from the resonance $^7$Be${^\ast_X}^{\ast{\rm a}}$(2P) to $^7$Be${^\ast_X}^{\ast{\rm a}}$(2S) and $^7$Be$^\ast_X$(1S).

\subsubsection{$^7$Be($X^-$, $\gamma$)$^7$Be$_X$ Nonresonant Rate}\label{sec3a3}

The nonresonant rate for the reaction $^7$Be($X^-$, $\gamma$)$^7$Be$_X$ in the temperature region of $T_9=[10^{-3}, 1]$ has been calculated and fitted\cite{Kusakabe:2014moa} to be
\begin{equation}
N_{\rm A} \langle \sigma v \rangle_{\rm NR} = 
3.86 \times 10^4~{\rm cm}^3 {\rm mol}^{-1} {\rm s}^{-1}  \left(1-0.194 T_9 \right) T_9^{-1/2}. \label{eq31}
\end{equation}
The nucleosynthesis for $^4$He and heavier nuclei as well as nuclear recombinations with $X^-$ proceed after the temperature of the universe decreases to $T_9<1$.  The reaction rates for higher temperatures $T_9>1$ are, therefore, not necessary in BBN calculations.  

Nonresonant cross sections have been calculated  with RADCAP taking into account the multiple components of partial waves for scattering states.  We show continuum wave functions at the CM energy $E=0.07$ MeV, which is the average energy corresponding to the temperature of the recombination of $^7$Be+$X^-$ for the case of $m_X=1000$ GeV, i.e., $E=3T/2$ with $T\sim 0.4 \times 10^9$ K.

The total cross section for the absorption of an unpolarized photon with frequency $\nu$ via an E1 transition from a bound state ($n$, $l$) to a continuum state ($E$) is given\cite{Gaunt1930,Karzas1961} by
\begin{eqnarray}
\sigma_{nl\rightarrow E}&=&\frac{16 \pi^2}{3} e_1^2 \mu k \nu \nonumber\\
&&\times \left[\frac{l+1}{2l+1} \left( \tau_{nl}^{E,~l+1}\right)^2 + \frac{l}{2l+1} \left( \tau_{nl}^{E,~l-1}\right)^2 \right],\nonumber\\
\label{eq32}
\end{eqnarray}
where
$k=\sqrt{\mathstrut 2\mu E}$ is the wave number, and
\begin{equation}
\tau_{nl}^{E,~l\pm1}=\int r^2 dr \psi_{E,~l\pm1}(r) r \psi_{nl}(r),
\label{eq33}
\end{equation}
is the dipole radial matrix element for the radius $r$, and wave functions are normalized as
\begin{equation}
\int r^2 dr \left|\psi_{nl}(r)\right|^2=1,
\label{eq34}
\end{equation}
and asymptotically at large $r$
\begin{equation}
\psi_{E,~l}(r)\sim \frac{\sin\left[kr -\eta \ln (2kr) -l\pi/2+ \sigma_l +\delta_l\right]}{kr},
\label{eq35}
\end{equation}
where $\eta$ is defined by
\begin{equation}
\eta=\frac{Z}{ka_{\rm B}}=\left( \frac{Z^2 \alpha^2 \mu}{2E} \right)^{1/2}.
\label{eq38}
\end{equation}
with $a_{\rm B}=1/(\mu \alpha)$ the Bohr radius,
$\sigma_l$ is the Coulomb phase shift, and
$\delta_l$ is the phase shift due to the difference in Coulomb potential between cases of the point charge and finite size nuclei.\cite{Burke2011}
  The important point is that we must use exact values of the reduced mass $\mu$ [Eq. (\ref{eq_add1})] and the effective charge of E1 transitions [Eq. (\ref{eq16})] compared to the case of hydrogen-like electronic ions.

The calculated cross sections are compared with those for the recombination of two point charges.  For a system of two point charges, wave functions of scattering and bound states, and the bound-free absorption cross section have been derived analytically.\cite{Karzas1961,Kusakabe:2014moa}  Again, appropriate values of $\mu$ and $e_1$ are to be used in the analytic formulae for the present system.

\begin{figure}
\begin{center}
\includegraphics[width=0.6\textwidth]{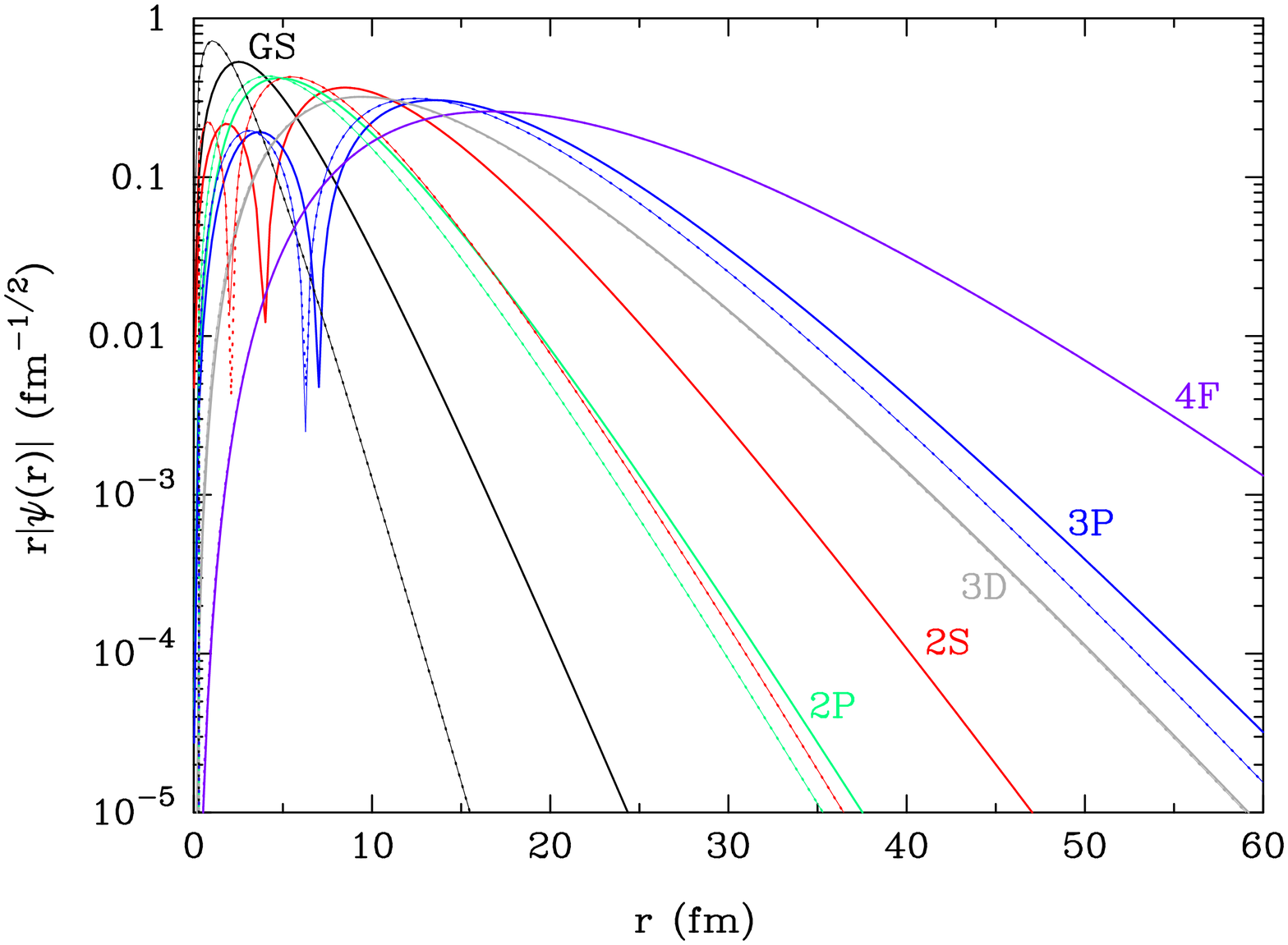}\\
\includegraphics[width=0.6\textwidth]{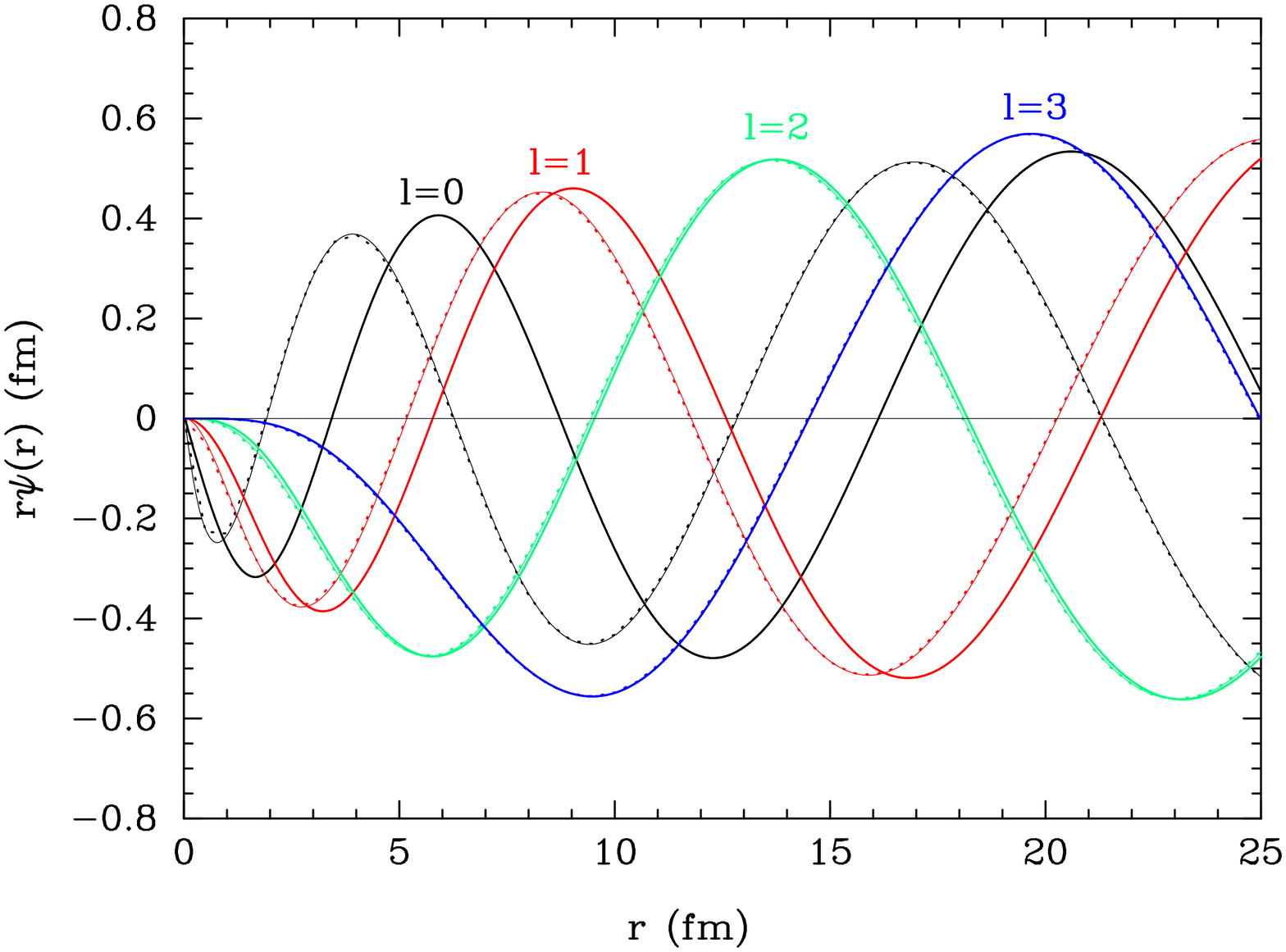}\\
\includegraphics[width=0.6\textwidth]{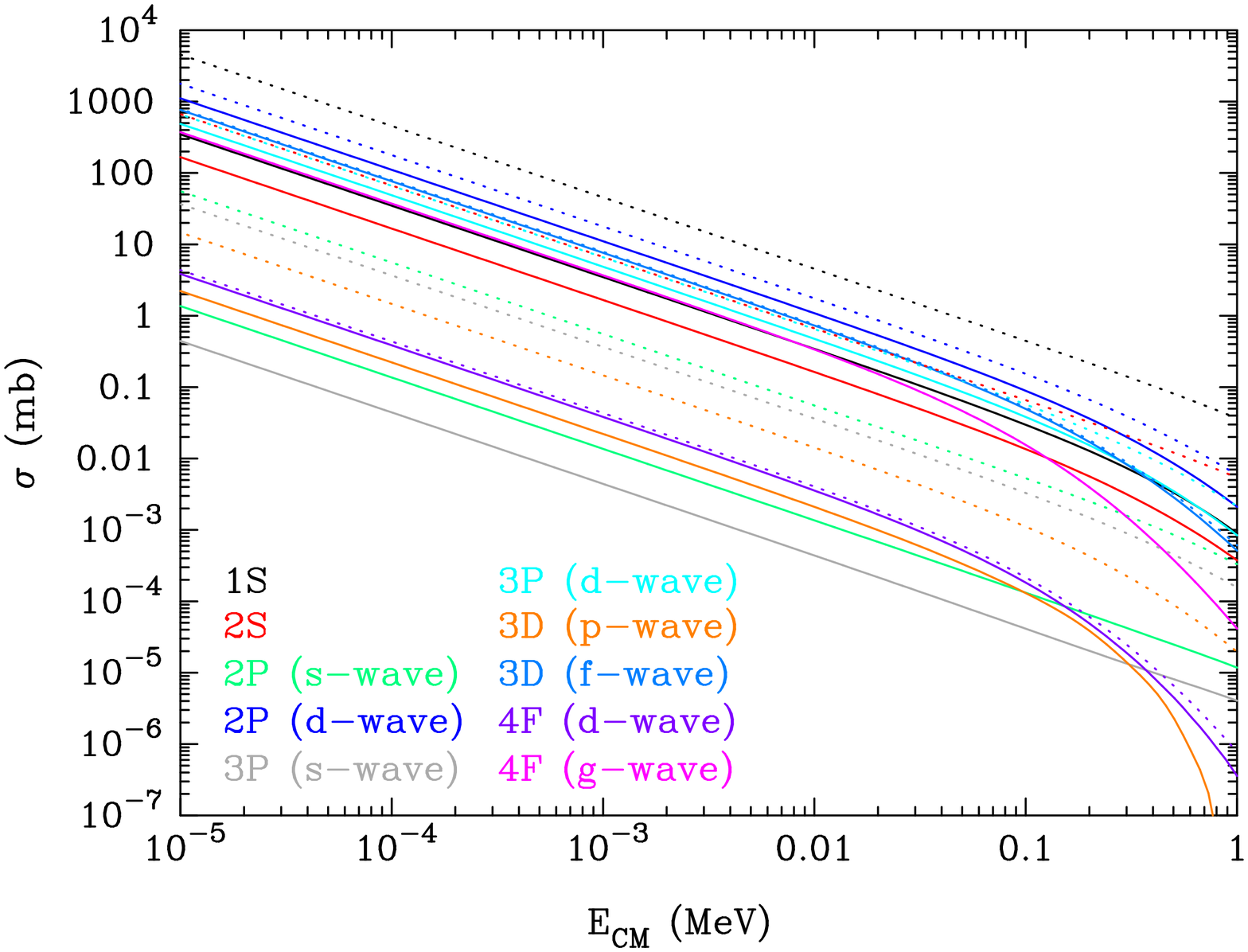}
\end{center}
\caption{Bound-state wave functions (upper panel) and continuum wave functions at $E=3T/2=0.07$ MeV (middle panel) for  the $^7$Be+$X^-$ system as a function of radius $r$ for the case of $m_X=1000$ GeV.  The bottom panel shows the recombination cross section as a function of CM energy $E$.  The thick solid lines correspond to calculated results for the case of a finite size $^7$Be charge distribution.  The thin solid lines in the upper and middle panels show calculated results for the point charge $^7$Be.  The dotted lines in the bottom panel correspond to analytic formulae for hydrogen-like atomic states composed of two point charges.  This figure is reprinted from Ref. \protect \refcite{Kusakabe:2014moa}. \label{fig9}}
\end{figure}

Figure \ref{fig9} shows bound-state wave functions (upper panel) and continuum wave functions (middle panel) at $E=0.07$ MeV for the $^7$Be+$X^-$ system as a function of radius $r$ for the case of $m_X=1000$ GeV.   Thick solid lines correspond to calculated wave functions.  Thin solid lines show calculated wave functions for the case of point-charge nuclei, and dotted lines correspond to the analytic formula for point-charge nuclei.  In fact, dotted lines overlap thin solid lines almost completely and no difference is seen.
In the upper panel, wave functions for the bound GS (1S state), 2S, 2P, 3P, 3D, and 4F states are plotted.  The wave functions for the GS and 2S state in the finite charge distribution case (thick lines) significantly deviate from those of the point charge case (thin lines).  While a certain degrees of deviations are seen in wave functions of bound 2P and 3P states also, the wave functions of 3D and 4F states agree with those for the point charge case.  The scattering wave functions for the  $s$-, $p$-, $d$-, and $f$-waves are plotted in the middle panel.  The wave functions of the $l=0$ and $l=1$ states for the finite charge distribution case deviates from those of the point charge case.


The bottom panel shows the recombination cross section as a function of the energy $E$.  Solid lines correspond to the calculated results, while the dotted lines correspond to the analytic solution for the two point charges.  Partial cross sections for the following transitions are drawn:  scattering $p$-wave $\rightarrow$ bound 1S state (black lines); $p$-wave $\rightarrow$ 2S (red); $s$-wave $\rightarrow$ 2P (green); $d$-wave $\rightarrow$ 2P (blue); $s$-wave $\rightarrow$ 3P (gray);  $d$-wave $\rightarrow$ 3P (sky blue); $p$-wave $\rightarrow$ 3D (orange);  $f$-wave $\rightarrow$ 3D (cyan); $d$-wave $\rightarrow$ 4F (violet); and $g$-wave $\rightarrow$ 4F (magenta).  

Because of the large difference in the scattering $s$-wave function, the cross sections for transitions from an initial $s$-wave, i.e., $s$-wave $\rightarrow$ 2P and $s$-wave $\rightarrow$ 3P are much  smaller than those in the point charge case.  Partial cross sections for transitions from an initial $p$-wave to bound 1S, 2S and 3D states are also altered  by the finite-size charge distribution.  The cross sections for transitions to 1S and 2S states are affected additionally by differences in binding energies of the states between the finite- and point-charge cases.

An important characteristic of the $^7$Be+$X^-$ recombination has been found.:    In the limit of a heavy $X^-$ particle, i.e., $m_X\gtrsim 100$ GeV, the most important transition in the nonresonant recombination is the $d$-wave $\rightarrow$ 2P.  This is different from the case of the point charge model, where the transition $p$-wave $\rightarrow$ 1S is predominant (see dotted lines).  In the case of a finite size charge distribution, in addition to the main pathway of $d$-wave $\rightarrow$ 2P, cross sections for the transitions $f$-wave $\rightarrow$ 3D and $d$-wave $\rightarrow$ 3P are also larger than that for the GS formation.  

The latest rate\cite{Kusakabe:2014moa} is more than 6 times larger than the previous rate\cite{Bird:2007ge}.
This large difference is caused since the most important transition from the scattering $d$-wave to the bound 2P state, and many other transitions are taken into account in the latest calculation.

\subsection{$^7$Li}\label{sec3b}
\begin{figure}
\begin{center}
\includegraphics[width=0.8\textwidth]{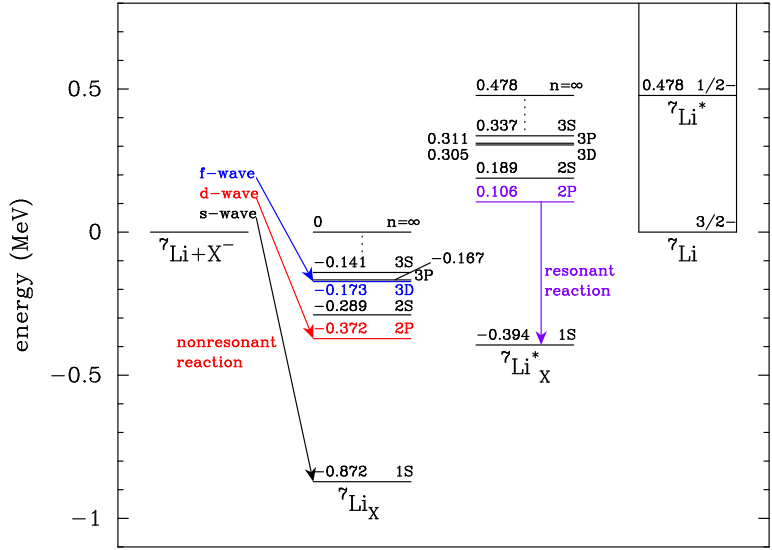}
\end{center}
\caption{Energy level diagram of $^7$Li$_X$ in the case of $m_X=1000$ GeV.  Red, black, and blue arrows show the three important transitions in the nonresonant recombination reaction of $^7$Li($X^-$, $\gamma$)$^7$Li$_X$, while a purple arrow shows the decay of the important resonance into a bound state of $^7$Li$^\ast_X$ in the resonant reaction.  \label{fig_new2}}
\end{figure}

Figure \ref{fig_new2} shows the energy level diagram of $^7$Li$_X$ in the case of $m_X=1000$ GeV.  The nonresonant rate of the recombination of $^7$Li and  $X^-$ is larger than the resonant rate.\cite{Kusakabe:2014moa}  Red, black, and blue arrows show the three important transitions in the nonresonant recombination reaction, while a purple arrow shows the decay of the important resonance into a bound state of $^7$Li$^\ast_X$ in the resonant reaction.


Similarly to the recombination of $^7$Be+$X^-$, the recombination can efficiently proceed via resonant reactions involving atomic states of ${^7{\rm Li}^\ast_X}^{\ast{\rm a}}$ composed of the first nuclear excited state $^7$Li$^\ast$[$\equiv ^7$Li$^\ast$(0.478 MeV, $1/2^-$)].  

The resonant rate for the reaction $^7$Li($X^-$, $\gamma$)$^7$Li$_X$ has been calculated\cite{Kusakabe:2014moa} to be
\begin{equation}
  N_{\rm A} \langle \sigma v \rangle_{\rm R} = 
1.68 \times 10^4~{\rm cm}^3 {\rm mol}^{-1} {\rm s}^{-1} T_9^{-3/2} \exp(-1.22/T_9). \label{eq46}
\end{equation}
The rate correspond to the atomic transition from the resonance $^7$Li${^\ast_X}^{\ast{\rm a}}$(2P) to $^7$Li$^\ast_X$(1S).

The thermal nonresonant rate is given\cite{Kusakabe:2014moa} by
\begin{equation}
N_{\rm A} \langle \sigma v \rangle_{\rm NR} = 
1.62 \times 10^4~{\rm cm}^3 {\rm mol}^{-1} {\rm s}^{-1}  \left(1-0.245 T_9 \right) T_9^{-1/2}. \label{eq53}
\end{equation}

This $^7$Li+$X^-$ system has the important characteristic that the transition $d$-wave $\rightarrow$ 2P is the most important for $m_X\gtrsim 100$ GeV, as in the $^7$Be+$X^-$ system.

\subsection{$^9$Be}\label{sec3c}

In the recombination of $^9$Be and $X^-$, there are no important resonances of atomic states composed of the nuclear excited states for $^9$Be$^\ast$.  Then, only the nonresonant rate is needed.
The nonresonant rate is then taken from Ref. \refcite{Kusakabe:2014moa}.
The transition, $d$-wave $\rightarrow$ 2P, is most important for $m_X\gtrsim 100$ GeV in this $^9$Be+$X^-$ system, also.

\subsection{$^4$He}\label{sec3d}

Atomic states composed of nuclear excited states $^4$He$_X^\ast$ are never important resonances in the recombination process.  The nonresonant rate is then taken from Ref. \refcite{Kusakabe:2014moa}. 

Because of the small amplitude of the Coulomb potential, the effect of the finite-size nuclear charge does not significantly affect the wave functions and cross sections.  As a result, even in the case of heavy $X^-$ particles ($m_X\gtrsim 100$ GeV), the rate for the recombination of $^4$He and $X^-$ is contributed by the transition $p$-wave $\rightarrow$ 1S most, similarly to the case of the Coulomb potential for point charges.

\subsection{Other nuclei}\label{sec3e}

For other nuclides, we approximately adopt the Bohr atom  formula\cite{Bethe1957} in the estimation of recombination rates.  We adopt cross sections in the limit that the CM kinetic energy, $E$, is much smaller than the binding energy, $E_{\rm B}$.  This is justified since the condition, $E=\mu v^2/2=3T/2 \ll E_{\rm B}$ with $v$ the relative velocity of a nucleus $A$ and $X^-$, always holds when the bound state formation is more efficient than its destruction.  The cross sections are given by
\begin{equation}
\sigma_{\rm rec}=\frac{2^9 \pi^2 e_1^2}{3\mathrm{e}^4} \frac{E_{\rm B}}{\mu^3 v^2},
\label{eq62}
\end{equation}
where
$\mathrm{e}=2.718$ is the base of the natural logarithm.  The thermal reaction rate is then given by
\begin{eqnarray}
N_{\rm A} \langle \sigma_{\rm rec} v \rangle &=& \frac{2^{19/2} \pi^{3/2} N_{\rm A}e_1^2}{3\mathrm{e}^4} \frac{E_{\rm B}}{\mu^{5/2} T^{1/2}}\nonumber\\
&=&1.37\times 10^4~{\rm cm}^3~{\rm s}^{-1} \frac{(e_1/e)^2 Q_9}{A^{5/2}T_9^{1/2}}\nonumber\\
&\equiv& C_1 T_9^{-1/2}~{\rm cm}^3~{\rm s}^{-1},
\label{eq63}
\end{eqnarray}
where
$Q_9=Q/{\rm MeV}$ is the $Q$-value in units of MeV, and we defined a rate coefficient $C_1=1.37\times 10^4(e_1/e)^2 Q_9/A^{5/2}$.  The $Q$-value for the recombination is equal to the binding energy of the $X$-nucleus $E_{\rm B}$.

Rates of radiative recombination of $A$ and $X^-$ and photoionization of $A_X$ are taken from Ref. \refcite{Kusakabe:2014moa}.

\section{RESONANT PROTON CAPTURE REACTIONS}\label{sec4}
Two important resonant reactions are
\begin{eqnarray}
^7{\rm Be}_X&+&p \rightarrow ^8{\rm B}_X^{\ast{\rm a}}(2\mathrm{P}) \rightarrow ^8{\rm B}_X+\gamma \nonumber\\
&&~~~[Q=m(^7{\rm Be}_X)+m(p)-m(^8{\rm B}_X) =0.64~{\rm MeV}] \nonumber\\
^8{\rm Be}_X&+&p \rightarrow ^9{\rm B}_X^{\ast{\rm a}}(2\mathrm{P}) \rightarrow ^9{\rm B}_X+\gamma \nonumber\\
&&~~~[Q=m(^8{\rm Be}_X)+m(p)-m(^9{\rm B}_X) =0.33~{\rm MeV}],
\label{eq9}
\end{eqnarray}
where (2P) indicates the atomic 2P state, and $m(A)$ and $m(A_X)$ are masses of nucleus $A$ and $X$-nucleus $A_X$, respectively.
The states $^8$B$_X^{\ast{\rm a}}$(2P) and $^9$B$_X^{\ast{\rm a}}$(2P) are the first atomic excited states.  The resonant reactions through the atomic excited states are important since they result in $^7$Be$_X$ destruction and  $^9$B$_X$ production.  The superscript $\ast{\rm a}$ indicates an atomic excited state, that  is different from a nuclear excited state indicated by a superscript $\ast$.
Resonant rates for these radiative capture reactions can be calculated as follows.

The thermal reaction rate for isolated and narrow resonances is given\cite{Angulo1999} by
\begin{eqnarray}
N_{\rm A} \langle \sigma v \rangle&=& N_{\rm A} \left(\frac{2\pi}{\mu}\right)^{3/2} \omega \gamma T^{-3/2} \exp\left(-E_{\rm r}/T \right) \nonumber\\
&=&1.5394\times 10^{11}~{\rm cm}^3 {\rm mol}^{-1} {\rm s}^{-1} A^{-3/2} \omega \gamma_{,{\rm MeV}} \nonumber\\
&&\times T_9^{-3/2} \exp(-11.605E_{\rm r,MeV}/T_9),
\label{eq11}
\end{eqnarray}
where
$N_{\rm A}$ is Avogadro's number,
$A$ is the reduced mass in  atomic mass units (amu) given by $A=A_1 A_2/(A_1+A_2)$ with $A_1$ and $A_2$ the masses of two interacting particles, 1 and 2, in amu,
$T_9=T/(10^9~{\rm K})$ is the temperature in units of $10^9$ K.
The parameter $\omega$ is a statistical factor defined by
\begin{equation}
\omega=(1+\delta_{12})\frac{(2J+1)}{(2I_1+1)(2I_2+1)},
\label{eq12}
\end{equation}
where
$I_i$ is the spin of the particle $i$,
$J$ is the spin of the resonance,
and $\delta_{12}$ is the Kronecker delta necessary to avoid a double counting of identical particles.  The quantity $\gamma$ is defined  by
\begin{equation}
\gamma \equiv \frac{\Gamma_{\rm i} \Gamma_{\rm f}}{\Gamma(E_{\rm r})},
\label{eq13}
\end{equation}
where $\Gamma_{\rm i}$ and $\Gamma_{\rm f}$ are the partial widths for the entrance and exit channels, respectively.  $\Gamma(E_{\rm r})$ is the total width for a resonance with resonance energy $E_{\rm r}$, $\gamma_{,{\rm MeV}}$ is the $\gamma$ factor in units of MeV, and $E_{\rm r,MeV}$ is the resonance energy in units of MeV.

When $\omega=1$  as in the reactions considered here, and the radiative decay widths of $^8$B$_X^\ast$ and $^9$B$_X^\ast$, $\Gamma_\gamma$, are 
 much smaller than those for proton emission (as assumed here), the thermal reaction rate is given by
\begin{eqnarray}
N_{\rm A} \langle \sigma v \rangle&=&1.5394\times 10^{11}~{\rm cm}^3 {\rm mol}^{-1} {\rm s}^{-1} A^{-3/2} \Gamma_{\gamma,{\rm MeV}} \nonumber\\
&&\times T_9^{-3/2} \exp(-11.605E_{\rm r,MeV}/T_9)\nonumber\\
&\equiv& C T_9^{-3/2} \exp(-11.605E_{\rm r,MeV}/T_9),
\label{eq14}
\end{eqnarray}
where
$\Gamma_{\gamma,{\rm MeV}}=\Gamma_\gamma/({\rm 1~MeV})$ is the radiative decay width in units of MeV, and
$C$ is a rate coefficient determined from $A$ and $\Gamma_\gamma$.

The rate for a spontaneous emission via an electric dipole (E1) transition is given\cite{Blatt} by
\begin{equation}
\Gamma_\gamma=\frac{16\pi}{9}~e_1^2~E_\gamma^3~\frac{1}{2I_{\rm i}+1} \sum_{M_{\rm
i},~M_{\rm f}} \left|\int r Y_{1\mu}(\hat{r}) \Psi_{\rm f}^\ast \Psi_{\rm i}~d{\bfr}\right|^2,
\label{eq15}
\end{equation}
where
$e_1$ is the E1 effective charge [Eq. (\ref{eq16})], 
$E_\gamma$ is the energy of the emitted photon, $I_{\rm i}$ is the angular momentum of the initial state, $M_{\rm i}$ and $M_{\rm f}$ are magnetic quantum numbers of initial and final states with the subscript $\mu=M_{\rm i}-M_{\rm f}$. $\Psi_{\rm i}$ and $\Psi_{\rm f}$ are wave functions of the initial and final states, respectively, and $Y_{1\mu}(\hat{r})$ is the dipole spherical surface harmonic.

In the present system of $^{8,9}$B+$X^-$, the effective charge is $e_1=e(Z_{\rm B}m_X-Z_Xm_{\rm B})/(m_{\rm B}+m_X) \approx Z_{\rm B} e =5e$ in the limit of $m_X \gg m_A$, where $Z_{\rm B}=5$ and $Z_X=-1$ are the charge numbers of $^{8,9}$B and the $X^-$, respectively.

Resonant rates for the proton capture reactions are adopted from Ref. \refcite{Kusakabe:2014moa}, and the nonresonant rates are taken from Ref.~\refcite{Kamimura:2008fx}.

\section{$^9$B\lowercase{e} PRODUCTION FROM $^7$L\lowercase{i}}\label{sec5}

The dominant reaction of $^9$Be production in the BBN model with the $X^-$ particle is $^7$Li$_X$($d$, $X^-$)$^9$Be.\cite{Kusakabe:2014moa}  Both resonant and nonresonant components can contribute to the reaction rate.  Realistic calculations for this reaction rate are not available yet.  The current rate is based on the assumption that the astrophysical $S$ factor for the reaction is taken from the existing data for $^7$Li($d$, $n\alpha$)$^4$He, i.e, $S=30$ MeV b\cite{Caughlan1988}.  
This reaction is the key reaction since a signature of the $X^-$ particle is left on the primordial $^9$Be abundance through this reaction.  This model can, therefore, be tested in the future by observations of $^9$Be abundances in MPSs.  Therefore, realistic calculations of this reaction rate with quantum many-body models are needed.

\begin{figure}
\begin{center}
\includegraphics[width=0.8\textwidth]{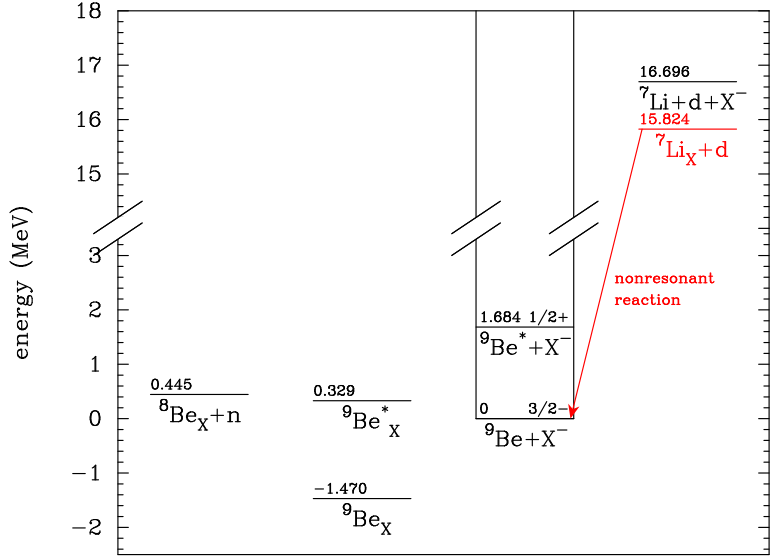}
\end{center}
\caption{Energy level diagram of the system of $^9$Be+$X^-$ in the case of $m_X=1000$ GeV.  Red arrow shows the most important reaction for the $^9$Be production, $^7$Li$_X$($d$, $X^-$)$^9$Be.  \label{fig_new3}}
\end{figure}

Figure \ref{fig_new3} shows the energy level diagram of $^9$Be$_X$ in the case of $m_X=1000$ GeV.  The red arrow shows the reaction for the $^9$Be production, $^7$Li$_X$($d$, $X^-$)$^9$Be.  Because of the relatively large abundances of $^7$Li$_X$ and $d$ (see Fig. \ref{fig31} below) and the large reaction rate, $^9$Be is produced significantly through this reaction.  It has been shown in a large network reaction calculation that other reactions cannot be responsible for significant $^9$Be production.\cite{Kusakabe:2007fv,Kusakabe:2014moa}

Although the resonant reaction $^8$Be$_X$($n$, $X^-$)$^9$Be through the atomic ground state of $^9$Be$^\ast$($1/2^+$, 1.684~MeV)$_X$, has been suggested for $^9$Be production,\cite{Pospelov:2007js} it was found that $^9$Be$^*$($1/2^+$, 1.684~MeV)$_X$ is located below the $^8$Be$_X$+$n$ threshold by a revised estimation using a more realistic nuclear charge radius\cite{Kamimura:2008fx}.  Energy levels of $^8$Be$_X$, $^9$Be$_X$, and $^9$Be$^\ast_X$ in Fig. \ref{fig_new3} correspond to the best estimate\cite{Kamimura:2008fx}.  The state $^9$Be$^\ast_X$ is below the state of the $^8$Be$_X$+$n$, and does not operate as a resonance.  This reaction is, therefore, unimportant.  We note that even if the state $^9$Be$^\ast_X$ is barely higher than the separation channel, this reaction rate is suppressed by an extremely small Coulomb penetration factor as shown [Sec. 2.3.2 in Ref. \refcite{Kusakabe:2007fv}] for the resonance $^8$B$^\ast$($1^+$,0.770~MeV)$_X$ in the reaction $^7$Be$_X$($p$,$\gamma$)$^8$B$_X$\cite{Kusakabe:2007fu}.


\section{BBN REACTION NETWORK}\label{sec6}

  We adopted the Kawano reaction network code\cite{Kawano1992,Smith:1992yy}, and utilized a modified version\cite{Kusakabe:2014moa}.  The free $X^-$ particle and bound $X$-nuclei are encoded as new species, and reactions involving the $X^-$ particle are encoded as new reactions.
The $\beta$-decay rates of $X$-nuclei ($A_X$) are taken from Ref. \refcite{Kusakabe:2014moa}.  Thermonuclear reaction rates of $X$-nuclei are adopted from Refs. \refcite{Hamaguchi:2007mp,Kamimura:2008fx,Kusakabe:2014moa}.  
Two parameters are updated in this review.  First, the neutron lifetime is the central value of the Particle Data Group, $880.3 \pm 1.1$~s\cite{Agashe:2014kda} based upon improved measurements\cite{Serebrov:2004zf,Mathews:2004kc,Serebrov:2010sg}.  The baryon-to-photon ratio is $(6.037 \pm 0.077) \times 10^{-10}$,\cite{Ishida:2014wqa} corresponding to the baryon density determined from the Planck observation of the cosmic microwave background, $\Omega_\mathrm{m} h^2 =0.02205 \pm 0.00028$ for the base $\Lambda$CDM model (Planck+WP+highL+BAO)\cite{Ade:2013zuv}.

$^9$Be$_X$ production through $^8$Be$_X$, i.e., $^4$He$_X$($\alpha$, $\gamma$)$^8$Be$_X$($n$, $\gamma$)$^9$Be, depends significantly on the energy levels of $^8$Be$_X$ and $^9$Be$_X$,\cite{Pospelov:2007js,Kamimura:2008fx,Cyburt:2012kp} and precise calculations with a quantum four body model by another group is under way\cite{Kamimura2010}.  In this paper, we neglect that reaction series, and leave that discussion to a future work.  The reaction $^4$He$_X$($\alpha$, $\gamma$)$^8$Be$_X$ is thus not included, and the abundance of $^8$Be$_X$ is not shown in the figures below.

\section{Abundance Constraints}\label{sec7}

Observational constraints on the deuterium abundance are taken from the weighted mean value of D/H$=(2.53 \pm 0.04) \times 10^{-5}$\cite{Cooke:2013cba}.  When the central values of adopted reaction rates, the neutron lifetime, and the baryon-to-photon ratio are used, the calculated abundances in the SBBN model is out of the $2\sigma$ observational limit.  The $4\sigma$ range is then adopted.

Constraints on the primordial $^3$He abundance are taken from the mean value of Galactic HII regions measured through the 8.665~GHz hyperfine transition of $^3$He$^+$, i.e., $^3$He/H=$(1.9\pm 0.6)\times 10^{-5}$\cite{Bania:2002yj}.

Constraints on the $^4$He abundance are taken from observational values of metal-poor extragalactic HII regions, i.e, $Y_{\rm p}=0.2551\pm 0.0022$\cite{Izotov:2014fga}, and adopt its $4\sigma$ range since the $2\sigma$ range is inconsistent with the theoretical abundances in the SBBN model.

We take the observational constraint on the $^7$Li abundance from the central  value of log($^7$Li/H)$=-12+(2.199\pm 0.086)$ derived in the  3D NLTE model of 
Ref.~\refcite{Sbordone:2010zi}.  

  Detections of $^6$Li abundances of MPSs have been reported.\cite{smith93,smith98,Cayrel:1999kx,Nissen:1999iq,Asplund:2005yt,ino05,asp2008,gar2009,ste2010,ste2012}  The measured abundance of $^6$Li/H$\sim 6\times10^{-12}$,\cite{Asplund:2005yt} is $\sim$1000 times higher than the SBBN prediction, and is also significantly higher than the prediction by a standard Galactic cosmic-ray nucleosynthesis model\cite{pra2006,pra2012}.  In a subsequent detailed analyses, however, it was found that most of the previous $^6$Li absorption feature  could be attributed to a combination of  3D turbulence and nonlocal thermal equilibrium (NLTE) effects in the model atmosphere.\cite{Lind:2013iza}  We adopt the least stringent $2~\sigma$ (95\% C.L.) upper limit among all limits reported in Ref.~\refcite{Lind:2013iza}, i.e., $^6$Li/H=$(0.9\pm 4.3)\times 10^{-12}$ for the G64-12 (NLTE model with 5 free parameters).

Spectroscopic observations of $^9$Be in MPSs show that the $^9$Be abundance scales linearly with Fe abundances generally.\cite{boe1999,Primas:2000ee,Tan:2008md,Smiljanic:2009dt,Ito:2009uv,Rich:2009gj,smi2012,ito2013}  The linear trend is explained with Galactic cosmic-ray nucleosynthesis models.\cite{ree1970,men1971,ree1974,pra2012}  Primordial abundances of Be before the start of the cosmic-ray nucleosynthesis, however, may be found by future observations.  We adopt the strongest upper limit on the primordial Be abundance, log(Be/H)$<-13.8$ derived from observations of carbon-enhanced MPS BD+44$^\circ$493 of an iron abundance [Fe/H]$=-3.8$ \footnote{[A/B]$=\log(n_{\rm A}/n_{\rm B})-\log(n_{\rm A}/n_{\rm B})_\odot$, where $n_i$ is the number density of $i$ and the subscript $\odot$ indicates the solar value, for elements A and B.} with Subaru/HDS\cite{Ito:2009uv,ito2013}.

\section{RESULTS}\label{sec8}
We show calculated results of BBN for $m_X=1000$ GeV.  See Ref. \refcite{Kusakabe:2014moa} for results for various mass values.  First, we analyze the time evolution for abundances of normal and $X$-nuclei.  Then we  update constraints on the parameters characterizing  the $X^-$ particle.

The two free parameters in this BBN calculation are the ratio of number abundance of the $X^-$ particles to the  total baryon density, $Y_X=n_X/n_{\rm b}$, and the decay lifetime of the $X^-$ particle, $\tau_X$.  The lifetime is assumed to be much smaller than the age of the present universe, i.e., $\ll14$ Gyr\cite{Hinshaw:2012fq}.  The primordial $X$-particles from the early universe are thus by now, long  extinct.  

As for the fate of $X$-nuclei, it is assumed that the total kinetic energy of products generated from  the decay of $X^-$ is large enough so that all $X$-nuclei can decay into normal nuclei plus the decay products of $X^-$.  The $X$ particle is detached from $X$-nuclei with its rate given by the $X^-$ decay rate.  The lifetime of $X$-nuclei is, therefore, given by the lifetime of the $X^-$ particle itself.

\subsection{Nucleosynthesis}
\begin{figure}
\begin{center}
\includegraphics[width=0.8\textwidth]{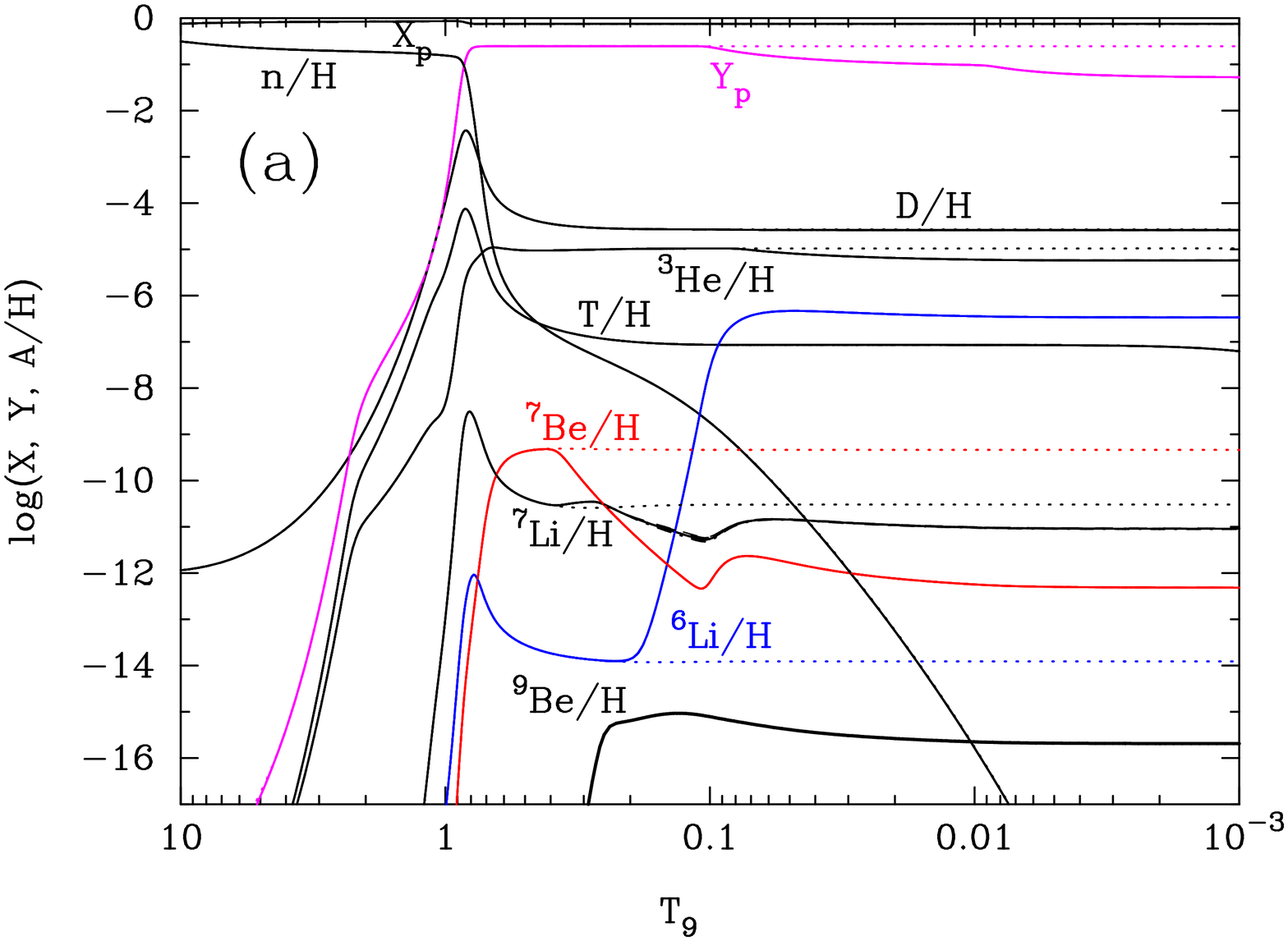}\\
\includegraphics[width=0.8\textwidth]{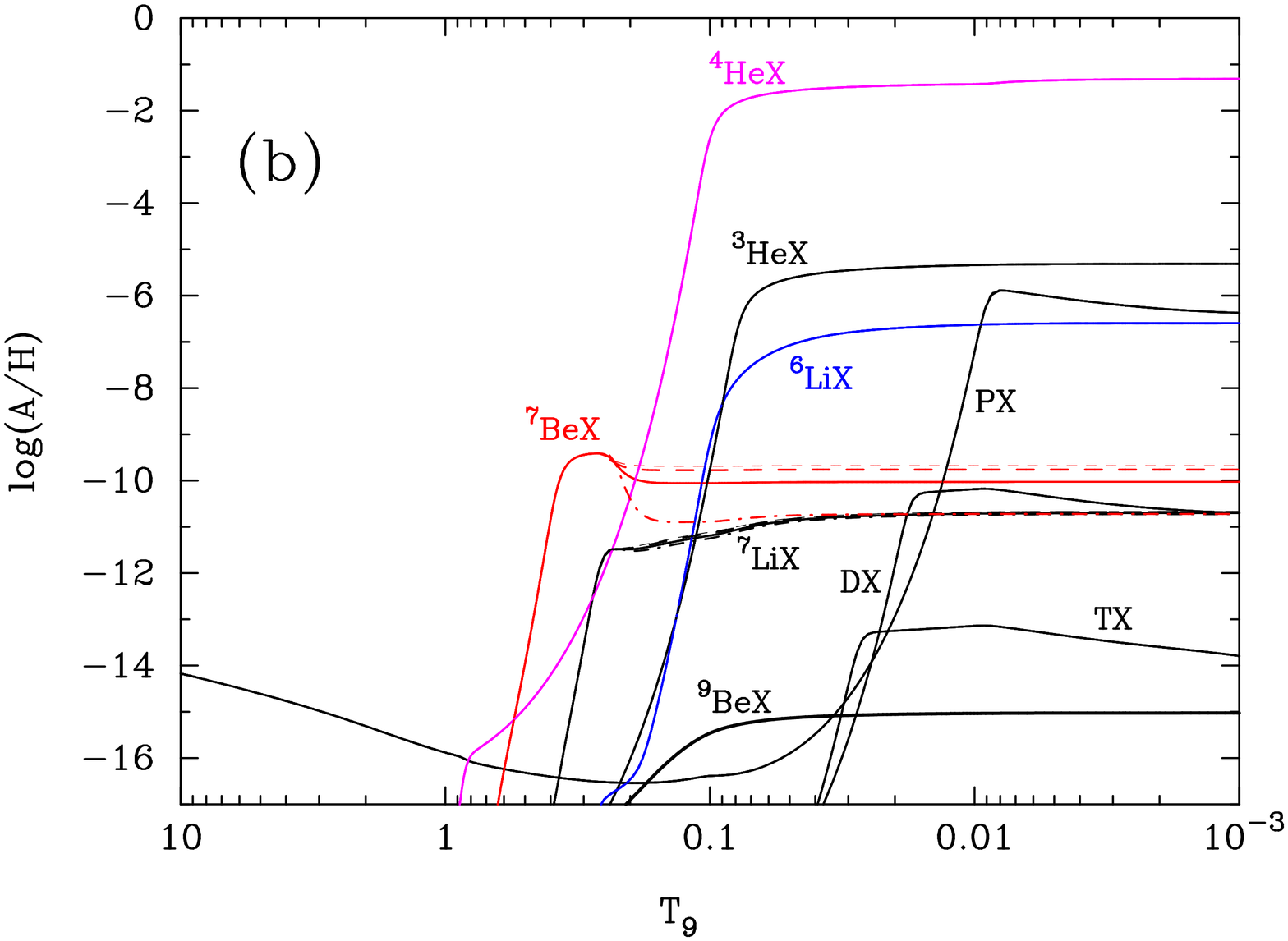}
\end{center}
\caption{Calculated abundances of normal nuclei (a) and $X$-nuclei (b) as a function of $T_9$.  $X_{\rm p}$ and $Y_{\rm p}$ are the mass fractions of $^1$H and $^4$He, in total baryonic matter, while the other curves correspond to number abundances with respect to that of hydrogen.  The abundance and  lifetime of the $X^-$ particle are taken to be $Y_X=n_X/n_b=0.05$ and $\tau_X=\infty$, respectively.  The dotted lines show the results of the SBBN model.  \label{fig31}}
\end{figure}

Figure \ref{fig31} shows the calculated abundances of normal nuclei (a) and $X$ nuclei (b) as a function of $T_9$.  Curves for $^1$H and $^4$He correspond to the mass fractions, $X_{\rm p}$ ($^1$H) and $Y_{\rm p}$ ($^4$He) in total baryonic matter, while the other curves correspond to number abundances with respect to that of hydrogen.  The dotted lines show the  result of the SBBN model.  The abundance and the lifetime of the $X^-$ particle are assumed to be $Y_X=0.05$ and $\tau_X=\infty$, respectively, for this example.


  Four lines are shown corresponding to different reaction rates for $^7$Be$_X$($p$, $\gamma$)$^8$B$_X$ and $^8$Be$_X$($p$, $\gamma$)$^9$B$_X$.  The two reactions are dominantly via resonant reactions whose rates are very sensitive to binding energies of $X$-nuclei.  The binding energies are affected by adopted nuclear charge distributions.  Therefore, we show results for the four different cases of charge distribution.  Three cases correspond to Gaussian (thick dashed lines), WS40 (solid lines), and well (dot-dashed lines) models, while one case corresponds to the previous calculation\cite{Kusakabe:2007fv} in which the reaction rate for $^7$Be$_X$($p$, $\gamma$)$^8$B$_X$ derived with a quantum many-body model\cite{Kamimura:2008fx} was adopted.  The amount of $^7$Be$_X$ destruction varies significantly when the nuclear charge distributions are changed.  The result of our Gaussian charge distribution model (thick dashed line) is close to that of the quantum many-body model. The differences in the curves for $^7$Be$_X$ thus indicate the effect of uncertainties in the charge density, which are estimated from measurements of RMS charge radii.

Early in the BBN epoch  ($10 \gtrsim T_9\gtrsim 1$), $p_X$ is the only $X$-nuclide with an abundance larger than $A_X/$H$>10^{-17}$.  Its abundance is the equilibrium value determined by the balance between  the recombination of $p$ and $X^-$ and the photoionization of $p_X$.  When the temperature decreases to $T_9\lesssim 1$, $^4$He is produced as in SBBN (panel a).  Simultaneously, the abundance of $^4$He$_X$ increases through the recombination of $^4$He and $X^-$ particles (panel b).  As the temperature decreases further, the recombination of nuclei with $X^-$ gradually proceeds in order of decreasing binding energies of $A_X$,  similar to the  recombination of nuclei with electrons.

$^7$Be first recombines with $X^-$ via the $^7$Be($X^-$, $\gamma$)$^7$Be$_X$ reaction at $T_9\lesssim 0.4$.  The produced $^7$Be$_X$ nuclei are then partially destroyed via the
$^7$Be$_X$($p$, $\gamma$)$^8$B$_X$ reaction.  In a later epoch at $T_9\lesssim 0.1$, the $^7$Be abundance increases mainly through the reaction $^4$He$_X$($^3$He, $X^-$)$^7$Be and somewhat less  through the reaction $^6$Li($p$, $\gamma$)$^7$Be.  The production rate via the former reaction is somewhat larger than that via the latter.

$^6$Li is produced through the reaction $^4$He$_X$($d$, $X^-$)$^6$Li at $T_9\sim 0.1$.  The abundance of $^6$Li$_X$ increases through the recombination reaction 
$^6$Li($X^-$, $\gamma$)$^6$Li$_X$.  Some  of the $^6$Li$_X$ nuclei are then destroyed through  proton capture via the $^6$Li$_X$($p$, $^3$He$\alpha$)$X^-$ 
reaction in the temperature range of $T_9\gtrsim 0.05$.

At $T_9\sim 0.3$--$0.2$ the $^7$Li abundance at first increases through the  neutron-induced through the two  reaction pathways of $^7$Be$_X$($n$, $p$$^7$Li)$X^-$ and $^7$Be$_X$($n$, $p$)$^7$Li$_X$($\gamma$, $X^-$)$^7$Li.   This is seen as a bump in the abundance curve.  This possible bump appears during the epoch when the recombination of $^7$Be has started but that of $^7$Li has not.
Then, the $^7$Li abundance decreases through recombination with $X^-$ at $T_9\lesssim 0.2$.  At $T_9\gtrsim 0.05$, the proton capture reaction $^7$Li$_X$($p$, 2$\alpha$)$X^-$ partly destroys $^7$Li nuclei produced via the recombination reaction.  Finally, $^7$Li is produced through the reaction $^4$He$_X$($t$, $X^-$)$^7$Li at $T_9\lesssim 0.1$.

$^9$Be is produced through the reaction $^7$Li$_X$($d$, $X^-$)$^9$Be at $T_9\sim 0.3$--$0.1$.  The recombination $^9$Be($X^-$, $\gamma$)$^9$Be$_X$ reaction enhances the abundance of $^9$Be$_X $at $T_9\sim 0.2$--$0.1$.  When the  proton-capture reaction $^9$Be$_X$($p$, $^6$Li)$^4$He$_X$ is operative at $T_9\gtrsim 0.1$,
 it decreases the abundance of $^9$Be$_X$.

Figure \ref{fig31} is an updated version of Fig. 31 of Ref. \refcite{Kusakabe:2014moa}.  Although differences between the figures are minor, the smaller baryon-to-photon ratio adopted in this review leads to a decrease in the $^7$Be/H abundance by $\sim 6$ \% in SBBN.  As a result, the $^7$Be abundance in the BBN model with $X^-$ is also smaller.  In addition, the $^9$Be/H abundance in the BBN model with $X^-$ is higher than that of the previous calculation.  The smaller baryon-to-photon ratio leads to larger abundances of $^7$Li and D.  Since the $^9$Be nuclei are produced from $^7$Li and D with a catalysis $X^-$, the final $^9$Be abundance
is larger.

\subsection{Constraints on the $X^-$ Particle}
\begin{figure}
\begin{center}
\includegraphics[width=0.8\textwidth]{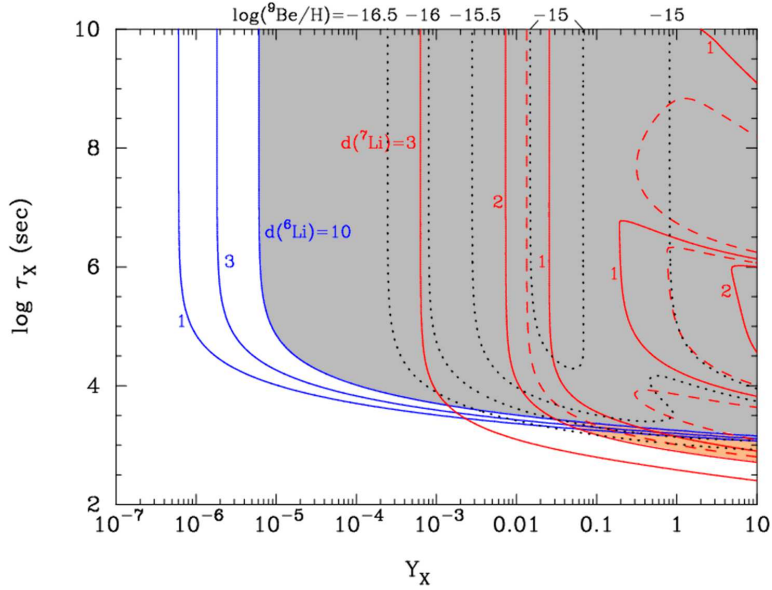}
\end{center}
\caption{Contours of constant lithium abundances relative to the observed values in MPSs, i.e., $d$($^6$Li)=$^6$Li$^{\rm Cal}$/$^6$Li$^{\rm Obs}$ (blue lines) and $d$($^7$Li)=$^7$Li$^{\rm Cal}$/$^7$Li$^{\rm Obs}$ (red lines).  The adopted observational constraint on the $^7$Li abundance is the central  value of log($^7$Li/H)$=-12+(2.199\pm 0.086)$ derived in the  3D NLTE model of Ref.~\protect\refcite{Sbordone:2010zi}.  The  $^6$Li constraint is taken from the $2 \sigma$  upper limit for the G64-12 [NLTE model with 5 parameters; Ref. \protect\refcite{Lind:2013iza}], of $^6$Li/H=$(0.9\pm 4.3)\times 10^{-12}$.  Dashed lines around the line of $d$($^{7}$Li)=1 correspond to the $2 \sigma$  uncertainty in the observational constraint.  The gray region located to the  right from the contours of $d$($^6$Li)=10 or the 2 sigma lower limit, $d$($^7$Li)=0.67, is  excluded by the overproduction of $^6$Li and underproduction of $^7$Li, respectively.  The orange region is the interesting parameter region in which a significant reduction in $^7$Li is realized without an overproduction of $^6$Li.  Dotted lines are contours of the abundance ratio of $^9$Be/H predicted when the unknown rate for the reaction $^7$Li$_X$($d$, $X^-$)$^9$Be is adopted as described in the text.  \label{fig32}}
\end{figure}

Figure \ref{fig32} shows contours of calculated final lithium abundances.  These are normalized to the values observed in MPSs, i.e., $d$($^6$Li)=$^6$Li$^{\rm Cal}$/$^6$Li$^{\rm Obs}$ (blue lines) and $d$($^7$Li)=$^7$Li$^{\rm Cal}$/$^7$Li$^{\rm Obs}$ (red lines).  It has been shown \cite{Kusakabe:2007fv,Kusakabe:2010cb} that concordance with the observational constraints on D, $^3$He, and $^4$He is maintained in the parameter region of $^7$Li reduction.  Therefore, constraints on only Li isotopes are seen.

The final $^7$Li abundance is a sum of the abundances of $^7$Li and $^7$Be produced in BBN.  This is because  $^7$Be is converted to $^7$Li via the electron capture at a later epoch.  The dashed lines around the line of $d$($^{7}$Li)=1 correspond to the 2$\sigma$ uncertainty in the observational constraint.  The gray region  located to the  right of the contours for  $d$($^6$Li)=10 and/or the 2 sigma lower limit, $d$($^7$Li)=0.67, are excluded by the overproduction of $^6$Li and underproduction of $^7$Li, respectively.  The orange region is the interesting parameter region in which a significant $^7$Li reduction occurs without an overproduction of $^6$Li.  Dotted lines are contours of the calculated abundance ratio of $^9$Be/H.


The excluded gray region corresponds to $Y_X \gtrsim 10^{-5}$ in the limit of a long $X^-$-particle life time $\tau_X \gtrsim 10^5$ s.  This region is determined from the overproduction of $^6$Li.  The orange region corresponding to  a solution to the $^{7}$Li problem is located at $Y_X \gtrsim 0.04$ and $\tau_X \sim (0.6$--$3) \times 10^3$ s.  Within this region, the primordial $^9$Be abundance is predicted to be $^9$Be/H$\lesssim 3\times 10^{-16}$.  This is much larger than the SBBN value\cite{Coc:2011az} of $9.60\times 10^{-19}$.  Since the abundances of D, $^3$He, and $^4$He are not significantly altered, the adopted constraints on their primordial abundances are satisfied in this region.

We note that the contour of $d$($^7$Li)=3 moved left significantly with respect to that in Fig. 32 of Ref. \refcite{Kusakabe:2014moa}.  This is because the abundance $^7$Li/H in SBBN is larger than that in the previous calculation by $\sim 6$ \% due to the smaller baryon-to-photon ratio adopted in this review.  Since abundances of $^7$Li and D are larger than the previous results, the abundance of $^9$Be produced via $^7$Li$_X$($d$, $X^-$)$^9$Be is also larger.  Contours of $^9$Be/H are, therefore, slightly changed by the larger calculated $^9$Be/H abundances.

\begin{figure}
\begin{center}
\includegraphics[width=0.8\textwidth]{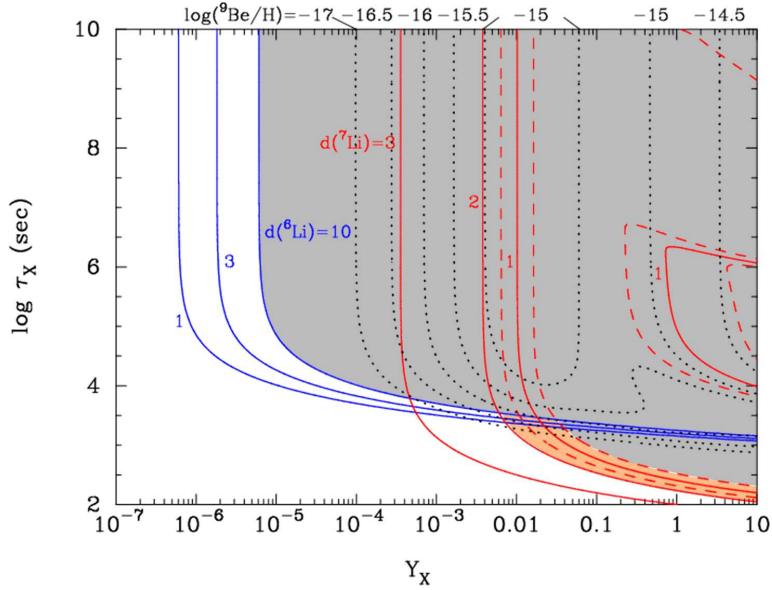}
\end{center}
\caption{Same as in Fig. \ref{fig32},  but the charged-current decay of $^7$Be$_X\rightarrow ^7$Li+$X^0$ is also included.  \label{fig33}}
\end{figure}

Figure \ref{fig33} shows the same contours for calculated abundances of $^{6,7}$Li and $^9$Be as in Fig. \ref{fig32}.  In this case  the instantaneous charged-current decay of $^7$Be$_X\rightarrow ^7$Li+$X^0$\cite{Jittoh:2007fr,Jittoh:2008eq,Jittoh:2010wh,Bird:2007ge} is also taken into account.  In this  case  the $X^-$ particle interacts not only via its charge but also a  weak interaction.\cite{Jittoh:2007fr,Jittoh:2008eq,Jittoh:2010wh}  $^7$Be$_X$ can then be converted to $^7$Li plus a neutral particle $X^0$.


The contours for the $^6$Li abundance are similar to those in Fig. \ref{fig32}.  On the other hand, the $^7$Li abundance is different from that in Fig. \ref{fig32} because of the different processes for $^7$Be destruction. Including the charged-current decay of $^7$Be$_X$, the destruction rate of $^7$Be in this model is the same as the recombination rate of $^7$Be itself.  In the model without the decay, the destruction rate requires  that $^7$Be$_X$ nuclei experience a proton capture reaction before being re-ionized to a $^7$Be+$X^-$ free state.
The different processes of $^7$Be$_X$ destruction, therefore, cause a difference in the efficiency for the final $^7$Li reduction.  In this model with the decay, the amount of $^7$Be destruction roughly scales as $Y_X$ unlike the model without the decay.

The excluded region is wider than in Fig. \ref{fig32}.
The region for the $^{7}$Li problem is at $Y_X \gtrsim 6\times 10^{-3}$ and $\tau_X \sim 10^2$--$4\times 10^3$ s.  In this region, the $^9$Be abundance is $^9$Be/H$\lesssim 3\times 10^{-16}$.

\section{Candidates of the $X^-$ particle}\label{sec9}
We consider possibilities that the long-lived CHAMP $X^-$ is a slepton, i.e., a supersymmetric partner of lepton, or Kaluza-Klein (KK) leptons, i.e., excited states of leptons realized in models for extra dimensions\cite{Feng:2003uy}.  In the former case, the gravitino $\grav$ is the lightest supersymmetric particle (LSP) and the sleptons $\slep$ are the next-to-LSP.  In the latter case, the KK graviton $G^1$ is the lightest KK particle (LKP) and the KK leptons $l^1$ are the next-to-LKP.  It is not assumed that the sleptons and the KK leptons can decay into sneutrinos and KK neutrinos via weak interaction, respectively.  Therefore, these models correspond to the case without the decay of $^7$Be$_X\rightarrow ^7$Li+$X^0$ (Fig. \ref{fig33}).

Especially, constraints on the long-lived stau ($\stau$) scenario have been studied using the thermal relic abundance.\cite{Asaka:2000zh}  The thermal annihilation rate of $X^-$ at a low temperature of $T\ll m_X$ roughly scales as
\begin{equation}
  \langle \sigma_{\rm ann} v \rangle \propto \frac{\alpha_{\rm em}^2}{m_X^2},
\label{eq_add2}
\end{equation}
where
$\alpha_{\rm em}$ is the fine structure constant.

The relic abundance of stau particles has been calculated in a specific case that the gravitino is the LSP and the stau is the next-to-LSP and the mass of the bino ($\tilde{B}$) is 1.1 times as large as the stau mass, i.e., $m_{\tilde{B}} =1.1 m_{\tilde{\tau}}$.\cite{Asaka:2000zh}  In the range of 10 GeV $\leq m_{\tilde{\tau}} \leq 10^4$ GeV, the relic abundance is roughly given by $n_{\tilde{\tau}}/s \sim 10^{-12}(m_{\tilde{\tau}}/1~{\rm TeV})$, where $s=(2\pi^2/45)g_{\ast{\rm S}}T^3$ is the entropy density with $g_{\ast{\rm S}}$ the numbers of massless degrees of freedom for entropy\cite{kolb1990}.  We use the number density of photon given by
\begin{equation}
  n_\gamma =\frac{2 \zeta(3)}{\pi^2} T^3,
\label{eq_add3}
\end{equation}
where $\zeta(3)=1.202$ is the Riemann zeta function.
Since the massless degrees of freedom is given by $g_{\ast{\rm S}}=3.91$ after the annihilation of electron and positron, the relic abundance\cite{Asaka:2000zh} corresponds to
\begin{eqnarray}
  Y_{\tilde{\tau}} &=&\frac{n_{\tilde{\tau}}}{s} \frac{s}{n_\gamma} \frac{1}{\eta} \nonumber \\
  &\sim& 0.01 \left( \frac{m_{\tilde{\tau}}}{1~{\rm TeV}} \right) \left( \frac{\eta}{6 \times 10^{-10}} \right)^{-1}.
\label{eq_add4}
\end{eqnarray}

From BBN calculations taking into account the effects of $X^-$ particles, it can be deduced that the abundance and lifetime of the $X^-$ should be $Y_X \gtrsim 0.04$ and $\tau_X \sim$ (0.6--3)$\times 10^3$ s for a significant reduction of the primordial $^7$Li abundance.  The thermal relic abundance\cite{Asaka:2000zh} is then consistent with the parameter region for the $^7$Li reduction if the stau mass is $m_{\tilde{\tau}} =\mathcal{O}$(1) TeV.

The decay rates of the NLSP sleptons are given\cite{Feng:2003uy} by
\begin{equation}
  \Gamma(\slep \rightarrow l\grav) = \frac{1}{48 \pi M_\ast^2}
  \frac{ m_{\tilde{l}}^5 }{ m_{\tilde{G}}^2 }
  \left( 1- \frac{m_{\tilde{G}}^2}{m_{\tilde{l}}^2} \right)^4,
\label{eq_add5}
\end{equation}
where
$M_\ast =M_{\rm Pl}/\sqrt{8\pi}=2.44 \times 10^{18}$ GeV is the reduced Planck mass with $M_{\rm Pl} =1.22 \times 10^{19}$ GeV the Planck mass, and
$m_i$ is the mass of particle $i$.  In the limit of $(m_{\tilde{G}} /m_{\tilde{l}})^2 \ll 1$, the lifetime is
\begin{equation}
  \tau(\slep \rightarrow l\grav) \sim 2 \times 10^3~{\rm s} \left( \frac{3~{\rm TeV}}{m_{\tilde{l}}} \right)^5 \left( \frac{m_{\tilde{G}}}{1~{\rm TeV}} \right)^2.
\label{eq_add6}
\end{equation}
Then, if the masses of sleptons and the gravitino are of order TeV, the lifetime is consistent with the interesting parameter region for $^7$Li reduction.

We must also consider effects of nonthermal nucleosynthesis and gravitino production triggered by slepton decay.  The constraints from electromagnetic energy injection and the gravitino energy density strongly limit the parameter space [Fig. 2 in Ref. \refcite{Asaka:2000zh}].  It is, however, found that a very narrow region exists near the maximum allowed gravitino mass of $\sim 1$ TeV.  In the decay of a selectron or smuon, generations of hadrons can be neglected.\cite{Feng:2003uy}  Therefore, there is no constraint from hadronic energy injection, and the parameter region for the $^7$Li problem remains unchanged.  On the other hand, the stau has a certain energy fraction of hadrons generated at the decay.  Hadronic energy injection is heavily constrained (e.g., Figs. 41 and 43 in Ref. \refcite{Kawasaki:2004qu}) as $\epsilon_{\rm had} n_X/s \lesssim 10^{-13}$ GeV with $\epsilon_{\rm had}$ the average initial hadronic energy generated in one $X$ decay\cite{Feng:2003uy}.  Even for the case of a small branching ratio to hadronic energy $\epsilon_{\rm had}/m_X \gtrsim 10^{-5}$\cite{Feng:2003uy}, a non-negligible constraint can be deduced $n_X/s \lesssim 10^{-11} (1~{\rm TeV}/m_X)$.  This constraint excludes the interesting parameter region in the stau NLSP scenario.

Because of the similarity of the interaction strength in the cases of supersymmetric and extra dimension models, the thermal relic abundances and the decay rates of sleptons and KK leptons are similar [cf. Ref. \refcite{Feng:2003uy}].  Therefore, depending on the masses of KK leptons and KK graviton, a solution to the Li problem exists for the cases of KK electron and muon, while it does not exists for the case of the KK tau.

\section{SUMMARY}\label{sec10}

We briefly reviewed the important reactions in the BBN model including a long-lived negatively charged massive particle, i.e., the $X^-$.  The mass of the $X^-$ is assumed to be much larger than the nucleon mass.  This model is worth studying since it can explain the observed $^7$Li abundances of MPSs which are smaller than that predicted in the SBBN model.  In addition, the primordial $^9$Be abundance can be larger than that of the SBBN model.  The existence of the $X^-$ particle can, therefore, be tested by future observations of the $^9$Be abundances in MPSs as well as collider searches for the $X^-$ particle.

After $^4$He synthesis during the BBN epoch, the $X^-$ particle starts recombination with background nuclei.  Since the $X^-$ particle is very heavy, i.e,. $m_X\gg$ GeV, the atomic size of the bound state $A_X$ is much smaller than those of hydrogen-like electronic ions.  As a result, the finite size charge distributions of nuclei cannot be neglected in describing the bound and scattering states.  Interesting characteristics then appear in the recombination rates of nuclei and $X^-$.
In the recombination of $^7$Li and $^{7,9}$Be with $X^-$, the dominant transition is the $d$-wave $\rightarrow$ 2P for $m_X\gtrsim 100$ GeV.  This is different from the formation of hydrogen-like electronic ions described by point-charge distribution, where the dominant transition is the $p$-wave $\rightarrow$ 1S.

In this model, the nuclide $^7$Be is destroyed via a recombination reaction with the $X^-$ followed by a radiative proton capture reaction, i.e., $^7$Be($X^-$, $\gamma$)$^7$Be$_X$($p$, $\gamma$)$^8$B$_X$.   Since the primordial $^7$Li abundance mainly originates from the abundance of $^7$Be produced during BBN, this $^7$Be destruction reduces the primordial $^7$Li abundance, and it can explain the observed abundances.  It should be noted that the resonant rate for radiative proton capture significantly depends on the adopted nuclear charge distribution.  In addition, much $^6$Li is produced via the recombination of $^4$He and $X^-$ followed by a deuteron capture reaction, i.e., $^4$He($X^-$, $\gamma$)$^4$He$_X$($d$, $X^-$)$^6$Li.

In addition, a new route for $^9$Be production, i.e, $^7$Li$_X$($d$, $X^-$)$^9$Be opens in this model.  The primordial $^9$Be abundance can be significantly enhanced by this reaction of $^7$Li$_X$.

We have performed detailed reaction network calculations of BBN, in which the formations of $X$-nuclei, the $\beta$-decays and nuclear reactions of $X$-nuclei, and their inverse reactions are included.   The most realistic constraints on the initial abundance and the lifetime of the $X^-$ particle were derived.  Parameter regions for the solution to the $^7$Li problem were identified, and the expected primordial $^9$Be abundances were calculated in the allowed parameter regions.

The derived abundance and lifetime of the $X^-$ particle can be realized in models of long-lived sleptons decaying into gravitinos, and KK leptons decaying into KK gravitons for $m_X\sim {\mathcal O}(1)$ TeV.  If the $X^-$ is the stau or the KK tau, however, their decays produce taus.  Since the decay of the tau involves a hadronic energy injection, it can easily change the light element abundances via nonthermal nuclear reactions.  We took into account this hadronic energy injection, and found that the parameter region for the reduction of $^7$Li does not exist in scenarios involving the stau and KK tau.  Long-lived selectrons, smuons, KK electrons and muons, on the other hand, do not induce significant hadronic energy injections.  In these cases, therefore, a very narrow parameter region exists for the reduction of $^7$Li.

\section*{Acknowledgments}

This work was supported by the National Research Foundation of Korea (Grant No. NRF-2014R1A2A2A05003548), and in part by Grants-in-Aid for Scientific Research of JSPS (26105517, 24340060).  Work at the University of Notre Dame supported by   
the U.S. Department of Energy under Nuclear Theory Grant DE-FG02-95-ER40934.

\end{document}